\documentclass[twocolumn, preprint2, twocolappendix]{aastex701}

\usepackage{amsmath}
\usepackage{url}
\usepackage[T1]{fontenc}
\usepackage{enumitem}

\pdfoutput=1

\received{October 23, 2025}

\revised{October 30, 2025; Revised December 5, 2025}

\accepted{December 6, 2025}

\submitjournal{ApJ}

\shorttitle{SPHEREx and NGC7023: PAHs, Aliphatics, and PAH Size}

\shortauthors{Boersma~et~al.}

\graphicspath{{./}{figures/}}

\begin{document}

\submitjournal{ApJ}

\title{SPHEREx: Aromatics, Aliphatics and PAH Size across the Iris Nebula}

\correspondingauthor{C.~Boersma}

\author[0000-0002-4836-217X]{C.~Boersma}
\affiliation{NASA Ames Research Center, MS 245-6, Moffett Field, CA 94035-1000, USA}
\email[show]{Christiaan.Boersma@nasa.gov}

\author[0000-0003-2552-3871]{A.~Maragkoudakis}
\affiliation{NASA Ames Research Center, MS 245-6, Moffett Field, CA 94035-1000, USA}
\email{Alexandros.Maragkoudakis@nasa.gov}

\author[0000-0002-6049-4079]{L.J.~Allamandola}
\affiliation{NASA Ames Research Center, MS 245-6, Moffett Field, CA 94035-1000, USA}
\email{allamandola@sbcglobal.net}

\author[0000-0002-1440-5362]{J.D.~Bregman}
\affiliation{NASA Ames Research Center, MS 245-6, Moffett Field, CA 94035-1000, USA}
\email{jdb849@gmail.com}

\author[0000-0002-8341-342X]{P.~Temi}
\affiliation{NASA Ames Research Center, MS 245-6, Moffett Field, CA 94035-1000, USA}
\email{Pasquale.Temi@nasa.gov}

\author[0000-0001-6035-3869]{V.J.~Esposito}
\affiliation{Schmid College of Science and Technology, Chapman University, Orange, CA 92866, USA}
\email{vesposito@chapman.edu}

\author[0000-0003-4716-8225]{R.C.~Fortenberry}
\affiliation{Department of Chemistry \& Biochemistry, University of Mississippi, University, MS 38677-1848, USA}
\email{r410@olemiss.edu}

\begin{abstract}

Observations by the SpectroPhotometer for the History of the Universe, Epoch of Reionization, and Ices Explorer (SPHEREx) are combined with counterpart Spitzer spectral map data to study the aromatic, aliphatic, and PAH size evolution across the northwest photo-dissociation region (PDR) of the Iris Nebula (NGC7023). The 3.3--3.4~\textmu m complex (I$_{\rm 3.3}$) and 11.2~\textmu m (I$_{\rm 11.2}$) PAH band strength are determined through direct integration. In addition, the former is decomposed into a 3.3 (I\textquotesingle$_{\rm 3.3}$) and 3.4~\textmu m (I\textquotesingle$_{\rm 3.4}$) sub-feature by fitting SPHEREx bandpass-integrated photometry using a modeled, highly sampled, multi-component spectrum. I$_{\rm 3.3}$, I$_{\rm 11.2}$, I\textquotesingle$_{\rm 3.3}$, and I\textquotesingle$_{\rm 3.4}$ all peak at the PDR. The NASA Ames PAH IR Spectroscopic Database is used to obtain the average number of carbon atoms ($\overline{\rm N_{C}}$) and small PAH fraction ($f_{\rm small}$) by fitting the isolated PAH component of the Spitzer segment; $70\lesssim\overline{\rm N_{C}}\lesssim76$ and $0.24\lesssim\text{f}_{\rm small}\lesssim0.36$. I\textquotesingle$_{\rm 3.4}$/I\textquotesingle$_{\rm 3.3}$, I$_{\rm 11.2}$/I$_{\rm 3.3}$, $\overline{\rm N_{C}}$, and $f_{\rm small}$ all show a demarcation that matches the large-scale morphology of the region. For I\textquotesingle$_{\rm 3.3}$ and I\textquotesingle$_{\rm 3.4}$ this is reflected by two distinct trends when plotted against each other, one associated with the dense, the other with the diffuse medium; $[N_{\rm H,ali}/N_{\rm H,aro}]_{\rm dense}$ = 0.42\textpm0.01 and $[N_{\rm H,ali}/N_{\rm H,aro}]_{\rm diffuse}$ = 0.10\textpm0.01. $\overline{\rm N_{C}}$ and $f_{\rm small}$  are tentatively correlated with I$_{\rm 11.2}$/I$_{\rm 3.3}$ (R=0.54\textpm0.05 and -0.45\textpm0.05, respectively). A wider variety of large(r) extended interstellar medium objects is required to tighten the correlations, turn them into quantitative calibrators for PAH size, and pin down the discrepancy of correlations with I\textquotesingle$_{\rm 3.3}$ involved.

\end{abstract}


\section{Introduction}
\label{sec:introduction}

The NASA Medium Class Explorer mission SpectroPhotometer for the History of the Universe, Epoch of Reionization, and Ices Explorer \citep[SPHEREx;][]{2020SPIE11443E..0IC} was successfully launched March 11, 2025. During its two-year mission it will be conducting the first near-infrared (IR), 0.75--5~\textmu m, all-sky spectro-imaging survey at a spatial resolution of 6\textquotedbl.2. Emission from polycyclic aromatic hydrocarbons (PAHs) dominates the IR emission of many astronomical objects, including interstellar medium (ISM) sources in the Milky Way, nearby galaxies, and even high-redshift galaxies \citep[][]{2008ARA&A..46..289T, 2023Nat......S}. The blended 3.3 and 3.4~\textmu m emission band complex is a prominent PAH feature covered by SPHEREx. Its potential to study the 3.3 and related 3.4~\textmu m components was recently explored by \citet{2025ApJ..992.....3Z} and \citet{2025arXiv250613863C}. Obtaining 3.3~\textmu m measurements of local interstellar medium (ISM) sources, especially those that are extended in nature, can provide valuable information on the evolution of the PAH aliphatic/aromatic ratio as well as PAH size \citep[e.g.,][]{1989ApJS...71..733A, 2012ApJ...760L..35L, 2017NewAR..77....1Y}. In particular with the 3.4/3.3 and 11.2/3.3~\textmu m PAH band strength ratios, where the former traces the aliphatic/aromatic ratio and the latter is considered the ultimate tracer for PAH size \citep[see e.g.,][]{2020MNRAS.494..642M}.

SPHEREx's spatial-spectral characteristics are well-paired with those of the Infrared Spectrograph \citep[IRS;][]{2004ApJS..154...18H} instrument onboard NASA's Spitzer Space Telescope \citep{2004ApJS..154....1W}. When used in its spectral mapping mode, Spitzer observations, which start at 5.2~\textmu m, can be extended down into the near-IR at comparable spatial and spectral resolution with SPHEREx data. The goal of this work is to perform such an analysis and study the aromatic, aliphatic, and PAH size evolution across the Iris Nebula (NGC7023), a reflection nebula (RN) located in Cepheus. NGC7023 is irradiated by HD200775, a B2Ve \citep[][]{1995ApJ...442..694R} massive spectroscopic binary \citep{2001ApJ...546..358M} Herbig~Be star. HD200775 is 430~pc away \citep{1997A&A...324L..33V}. Figure~\ref{fig:hst} depicts NGC7023's northwest photo-dissociation region (PDR). The PAH emission in this PDR has been extensively studied \citep[e.g.,][]{2007A&A...469..575B, 2011A&A...532A.128R, 2012A&A...542A..69P, 2013ApJ...769..117B, 2014ApJ...795..110B, 2015ApJ...806..121B, 2016ApJ...832...51B, 2016ApJ...819...65S, 2016A&A...590A..26C, 2018ApJ...858...67B, 2025A&A...700A.158M}. SPHEREx opens up a new opportunity to study the large-scale evolution of the aliphatic/aromatic ratio and PAH size across NGC7023.

\section{Observations and Data Reduction}
\label{sec:observations}

\begin{figure}
   \centering
    \includegraphics[width=\columnwidth]{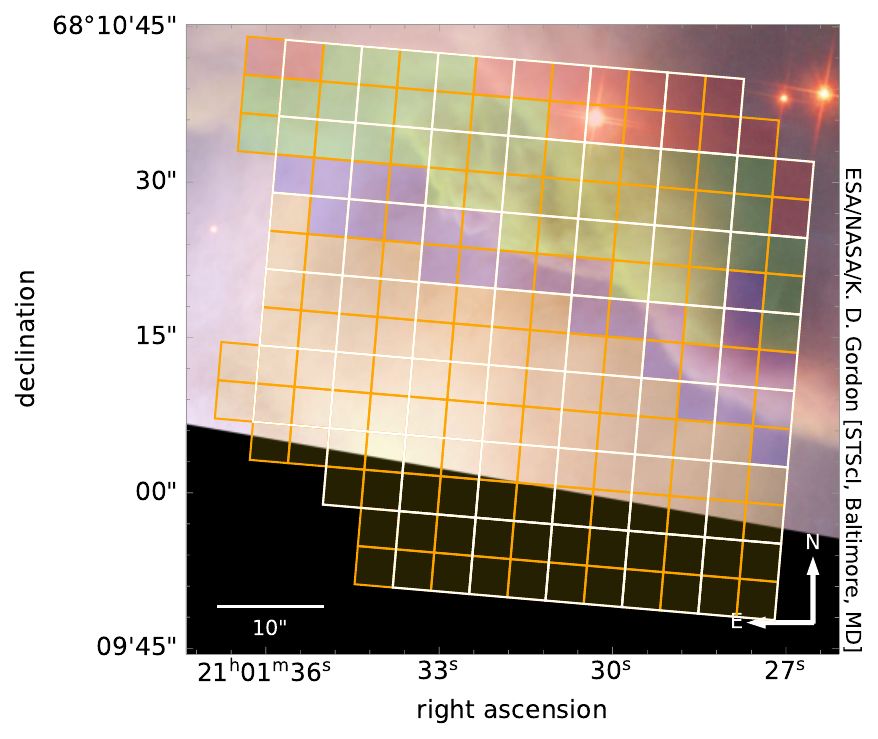}
    \caption{Composite Hubble Space Telescope ACS image of the northwest photo-dissociation region (PDR) in the reflection nebula NGC7023. Combined H$\alpha$ and infrared I-band data are shown in red, optical V-band data in green, and optical B-band data in blue. The irradiating star HD200775 is located some 43\textquotedbl\ from the center of the image toward the southeast, just outside the field of view. In orange Spitzer-IRS spectral map pixels are shown. Superimposed are four zones, colored red, green blue, and yellow, that were identified through hierarchical clustering of the Spitzer spectra, see \citet{2018ApJ...858...67B} for details. Blue marks the PDR-front, which forms the boundary between the dense medium toward the north from the diffuse medium in the south. Matching SPHEREx spectral extraction apertures are shown in white.}
    \label{fig:hst}
\end{figure}

Spitzer spectral map data on NGC7023 have been taken from \citet{2018ApJ...858...67B} and the reader is directed there for details on its reduction. The \texttt{astroquery} Irsa-interface was used to retrieve Multi-Extension FITS (MEF) images that have pixels overlapping the Spitzer spectral cube's field-of-view (FOV) from the SPHEREx Archive at the NASA/IPAC Infrared Science Archive \citep[IRSA;][]{10.26131/IRSA652}. These data are from the SPHEREx Quick Release 2 (QR2) and consist of observations from Survey~1 with Header Data Unit (HDU) creation dates spanning 2025-09-15 to 2025-09-19.

The calibrated surface brightness is provided in MJy/sr as 2040$\times$2040 pixel images with accompanying variances and predictions of the zodical light. Spectral coverage is from 0.75-5~\textmu m at a resolution $R$ ($\equiv\lambda/\Delta\lambda$) of 41-130.

\texttt{photutils} was used for spectral extraction, after subtracting out the predicted zodiacal light. To match the Spitzer with the SPHEREx observations, 2$\times$2 spatial pixels of the IRS map were combined\footnote[1]{The spectral map of NGC7023 from \citet{2018ApJ...858...67B} has already a 2$\times$2 spatial resampling applied. Therefore, each SPHEREx aperture matches 4$\times$4 native spatial Spitzer pixels.}, resulting in a 7$\times$7 spectral map with a pixel size of 7\textquotedbl.2$\times$7\textquotedbl.2, see Figure~\ref{fig:hst}. Associated uncertainties were determined by performing the same extraction on the variance maps and subsequently taking the square root.

Since SPHEREx is a spectro-photometer that utilizes a linear variable filter, each detector pixel has its own associated wavelength and bandwidth. A mapping is provided in the MEF files. Wavelength and width are assigned to each extracted spectral data point by averaging the values assigned to each SPHEREx pixel in the aperture, weighted with the overlap of the pixel with the aperture.

\texttt{pyphot} was used to predict Spitzer-IRAC channel~1 photometry from the extracted SPHEREx segment and IRAC channel 4 photometry from the 2$\times$2 resampled Spitzer segment to facilitate splicing. IRAC 1 and 4 Spitzer Enhanced Imaging Products (SEIP) Super Mosaics for NGC7023 were obtained from the Spitzer Heritage Archive (SHA) at IRSA. The mean mosaics were used in combination with \texttt{photutils} to retrieve matching photometric measurements for each of the SPHEREx apertures in Figure~\ref{fig:hst}.

\begin{figure}
    \centering
    \includegraphics[width=\columnwidth]{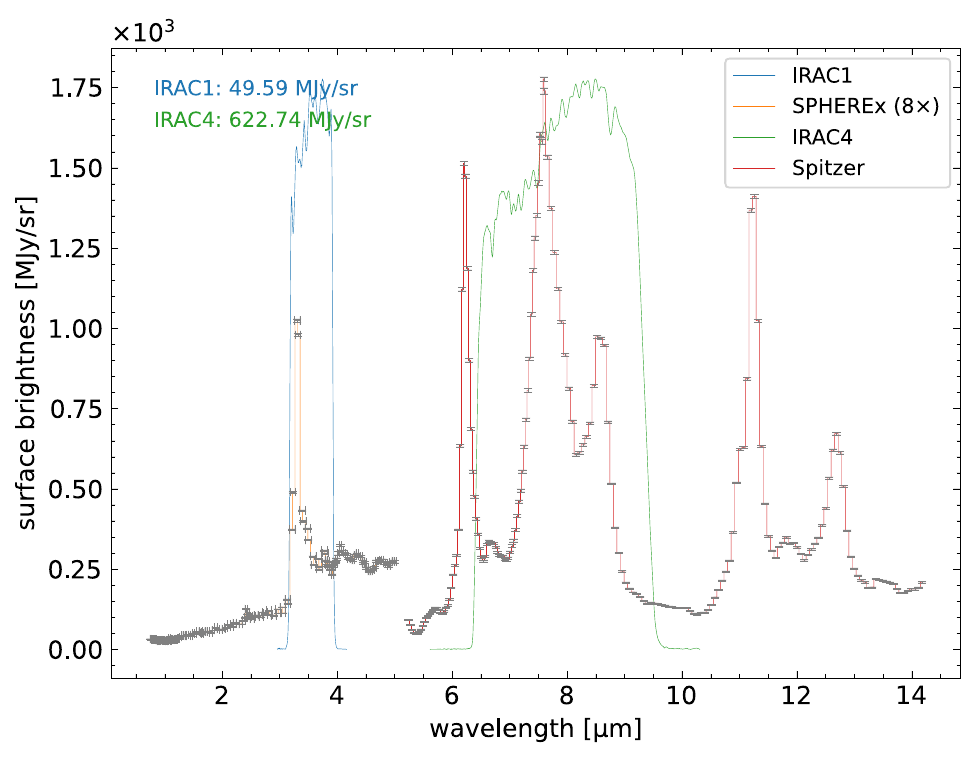}
    \caption{Extracted SPHEREx and resampled Spitzer segments for the center pixel of the 7$\times$7 constructed spectral cube, see Figure~\ref{fig:hst}. To highlight the SPHEREx segment, it is multiplied 8$\times$. Uncertainties are indicated with the gray vertical error bars. For the SPHEREx segment, horizontal error bars represent the bandpass of each spectro-photometric data point. IRAC channel 1 (blue) and 4 (green) bandpasses have been indicated as well as their predicted photometry.}
    \label{fig:extraction}
\end{figure}

Figure~\ref{fig:extraction} presents the extracted SPHEREx and Spitzer segments at the center pixel of the constructed spectral cube. The IRAC bandpasses are indicated as well as the predicted channel 1 and 4 photometry.

\section{Analysis}
\label{sec:analysis}

To study the aliphatic/aromatic ratio and PAH size evolution across the northwest PDR in NGC7023 through the 3.4/3.3 and 11.2/3.3~\textmu m PAH band strength ratios, respectively, the 3.3, 3.4, and 11.2~\textmu m PAH band strengths must be determined. This is done on extinction-corrected spectra, detailed in Appendix~\ref{app:extinction}, with measurements on the SPHEREx segment scaled to match the relative predicted and observed IRAC photometry, which is detailed in Appendix~\ref{app:splicing}. Two methods are used to determine these band strengths. 

\subsection{Direct Band Measurements}
\label{subsec:direct}

\begin{figure}
    \centering
    \includegraphics[width=\columnwidth]{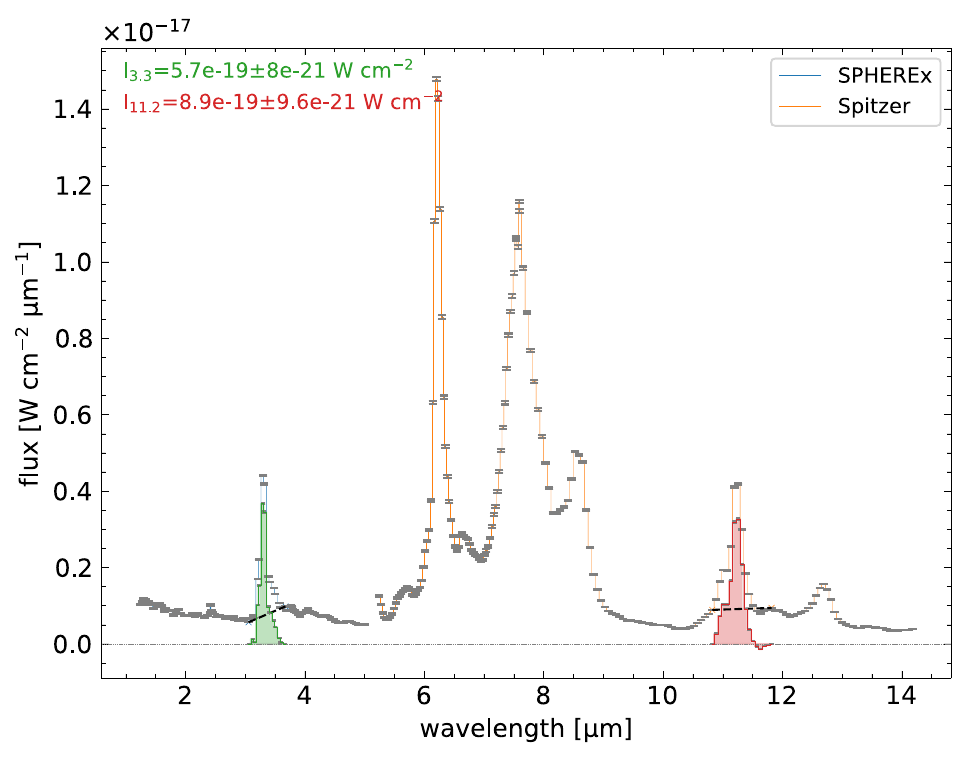}
    \caption{PAH band strength determination through direct integration. A straight baseline (dashed black) is drawn and subtracted to isolate the 3.3--3.4~\textmu m complex (green) and 11.2~\textmu m (red) PAH band. Uncertainties are indicated as gray error bars. The derived PAH band strengths and their uncertainty are indicated. See Section~\ref{subsec:direct} for details.}
    \label{fig:direct}
\end{figure}

The most straightforward way to determine PAH band strengths is through direct integration. Here, the PAH feature is isolated after subtracting a straight baseline with two anchor points. For the 3.3--3.4~\textmu m PAH band complex, these are 3.07 and 3.65~\textmu m and for the 11.2~\textmu m PAH band 10.8 and 11.8~\textmu m. Band strength uncertainties are determined by direct integration of the variance and taking a subsequent square root. Figure~\ref{fig:direct} demonstrates this approach for the center pixel of the 7$\times$7 constructed SPHEREx+Spitzer spectral cube of NGC7023.

\subsection{Spectro-photometric Modeled Band Determination}
\label{subsec:model}

\begin{figure*}
    \includegraphics[width=\linewidth]{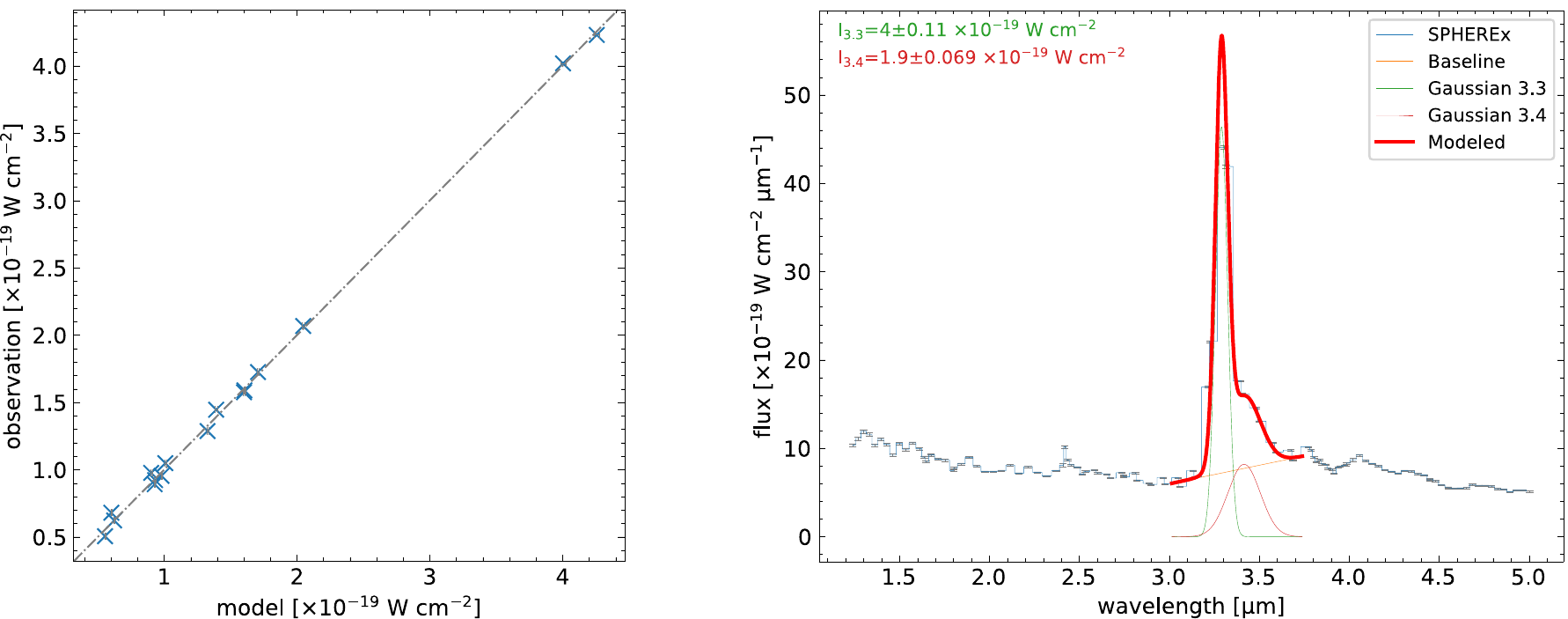}
    \caption{Left: Observed versus modeled SPHEREx bandpass-integrated photometry for the central pixel of the 7$\times$7 constructed spectral cube, see Fig.~\ref{fig:hst}. The dashed-dotted line shows the one-to-one line. Right: Modeled, highly sampled 3.3--3.4~\textmu m complex (red) overlain atop the observed SPHEREx segment (blue). Also shown are the individual 3.3 (green) and 3.4~\textmu m (maroon) component as well as the determined baseline (orange). Uncertainties are indicated as gray error bars. Derived band strengths and their uncertainties are given as well. See Section~\ref{subsec:model} for details.}
    \label{fig:model}
\end{figure*}

Alternatively, the SPHEREx observations can be fully treated as photometric data points. This has the advantage of being able to separate out the blended 3.3 and 3.4~\textmu m features, which is otherwise difficult to achieve given SPHEREx's spectral resolution. The 3.3 and 3.4~\textmu m features are commonly attributed to aromatic and aliphatic C-H stretches, respectively \citep[e.g.,][]{1989ApJS...71..733A, 2001ApJ...554..778L}.

In this approach the spectro-photometric SPHEREx observations are reproduced by fitting the bandpass-integrated photometry using the highly sampled spectral model that is described in Appendix~\ref{app:model}.

\texttt{astropy}'s modeling framework is used to implement the model, in combination with its \texttt{TRFLSQFitter} fitter. Figure~\ref{fig:model} presents the results for the SPHEREx segment of the central pixel of the spectral cube. The left panel of the figure compares the observed bandpass-integrated photometry versus its modeled counterpart. In the figure's right panel, the highly sampled spectrum is overlain atop the SPHEREx segment, indicating each of the spectral components and determined band strengths.

\section{Results and Discussion}
\label{sec:results}

\begin{figure*}
    \includegraphics[width=\textwidth]{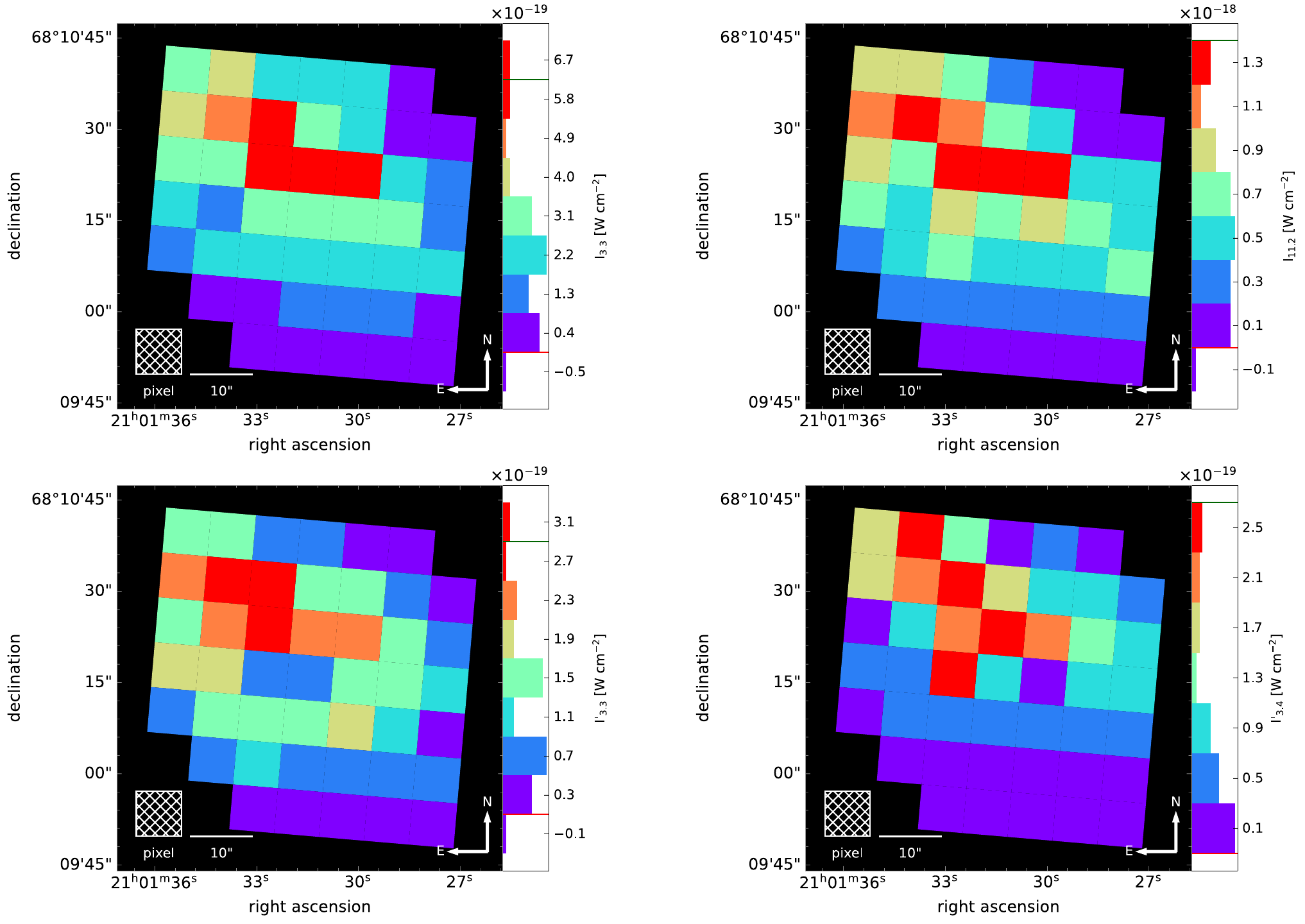}
    \caption{Maps of the northwest PDR in NGC7023 in the PAH 3.3--3.4~\textmu m complex, 3.3, 3.4 and 11.2~\textmu m bands. Top: 3.3--3.4~\textmu m complex and 11.2~\textmu m band strengths determined through direct measurement. See Section~\ref{subsec:direct} for details. Bottom: 3.3 and 3.4~\textmu m band strength determined using a multi-component photometric decomposition. See Section~\ref{subsec:model} for details. The color scales use a 4\% cutoff on the high and a 1\% cutoff on the low end, indicated by the green and red lines in the color bar, respectively. This suppresses outliers. See Section~\ref{sec:results} for details.}
    \label{fig:maps}
\end{figure*}

Maps of the determined 3.3, 3.4, and 11.2~\textmu m band strengths are presented in Figure~\ref{fig:maps}. The top two panels show those established through direct integration whereas the bottom two those from the spectro-photometric modeling. A distinction is made between either approach by decorating the latter with a quote `\textquotesingle', i.e., I$_{\rm 3.3}$ versus I\textquotesingle$_{\rm 3.3}$, respectively. Note that the maps of I$_{\rm 3.3}$, I\textquotesingle$_{\rm 3.3}$, and I\textquotesingle$_{\rm 3.4}$ have the scaling factors from Appendix~\ref{app:splicing} applied.

\subsection{The Spectra}
\label{subsec:spectra}

Focusing on the SPHEREx segment, Figures~\ref{fig:extraction}, \ref{fig:direct}, and \ref{fig:model} clearly show the 3.3--3.4~\textmu m PAH complex, which is made up of the blended 3.3 and 3.4~\textmu m sub-features. A jump is seen around 3~\textmu m that persists beyond the 3.3--3.4~\textmu m PAH complex up to $\sim$4.4~\textmu m with signs of substructure at 3.8, 4.0, and 4.4~\textmu m. This broad band emission has been attributed to a quasi PAH continuum, ascribed to 2-quanta combination bands and overtones of the strong PAH fundamental bands in the 5--15~\textmu m region \citep[][]{1989ApJS...71..733A, 2023ApJ...959...74B, 2023MolPhy2252936E}.

A stellar component, best visible in the $F_{\lambda}$-$\lambda$ spectra of Figures~\ref{fig:direct} and \ref{fig:model}, is also evident. It is particularly prominent in the northwest corner of the FOV, where an embedded young stellar object (YSO) is located, see Figure~\ref{fig:hst}.

\subsection{The Maps}
\label{subsec:maps}

The maps in Figure~\ref{fig:maps} show a clear demarcation between the dense and diffuse medium toward the north and south of the FOV, respectively, that matches the morphology apparent in Figure~\ref{fig:hst}. At the PDR, the emission peaks and gradually falls off when crossing into the diffuse medium and when nearing the embedded YSO located toward the northwest.

The morphologies of the 3.3, 3.4, and 11.2~\textmu m maps match up well. This is expected due to their common origin in C-H vibrational modes. The 11.2~\textmu m map mirrors that from \citet{2014ApJ...795..110B}, albeit at a coarser spatial resolution. There is no sign of the arc that can be discerned in some of the PAH band strength maps from \citeauthor{2014ApJ...795..110B} (\citeyear{2014ApJ...795..110B}; their Figure~5), and that is faintly visible in Figure~\ref{fig:hst}. \citet{2025A&A...700A.158M} report an offset between the peak of the aromatic I\textquotesingle$_{\rm 3.3}$ and aliphatic I\textquotesingle$_{\rm 3.4}$ emission in high spectral-spatial resolution James Webb Space Telescope \citep[JWST;][]{2006SSRv..123..485G} observations that is not resolved here.

\subsection{The 3.4/3.3 \texorpdfstring{\textmu}{u}m Band Strength Ratio}
\label{subsec:34_33}

\begin{figure*}
    \hfill\includegraphics[width=\textwidth]{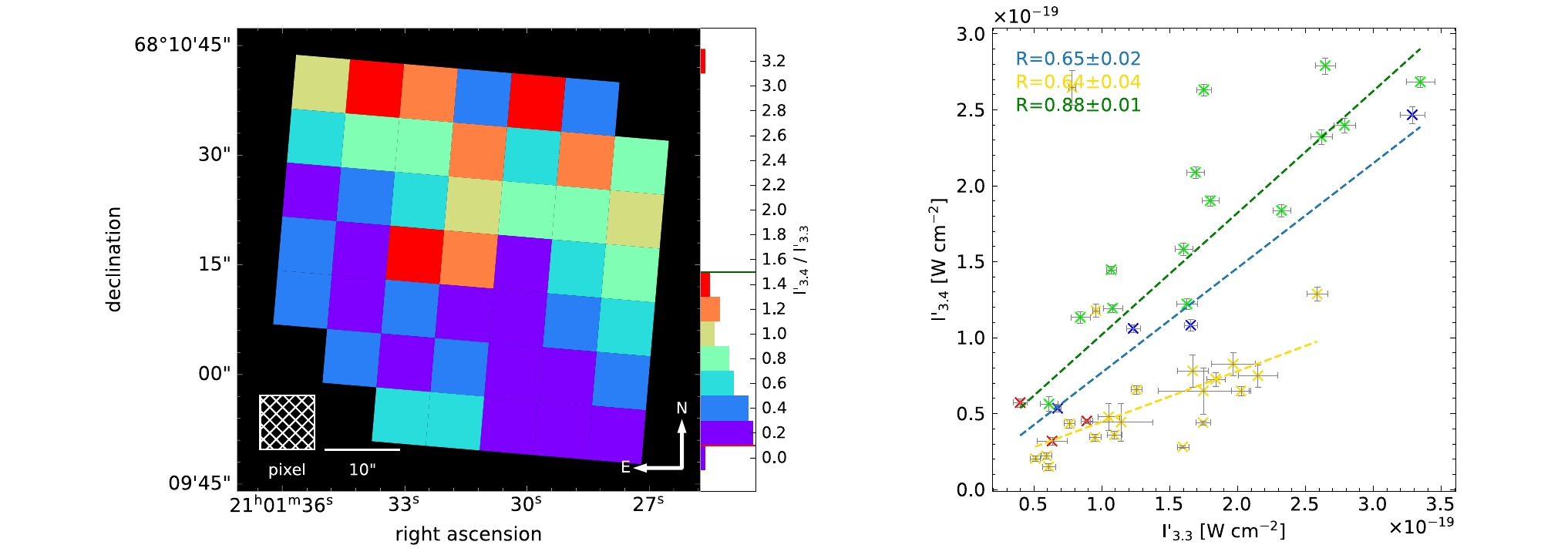}
    \caption{Left: Map of the northwest PDR in NGC7023 in the 3.4/3.3~\textmu m band strength ratio, derived using spectro-photometric modeling. The color scale uses a 4\% cutoff on the high and a 1\% cutoff on the low end, indicated by the green and red lines in the color bar, respectively. This suppresses outliers. Right: I\textquotesingle$_{\rm3.4}$ versus I\textquotesingle$_{\rm3.3}$. Points are color coded according their zone as assigned in Figure~\ref{fig:hst}. Uncertainties have been propagated, indicated as gray error bars, and points with a SNR less than three have been discarded. Given are the weighted Pearson correlation coefficients R for a straight line (dashed) associated with all data (blue), the diffuse medium (yellow), and the dense medium (green). See Section~\ref{subsec:34_33} for details.}
    \label{fig:aliphatics}
\end{figure*}

The left panel of Figure~\ref{fig:aliphatics} presents a map of the I\textquotesingle$_{\rm 3.4}$/I\textquotesingle$_{\rm 3.3}$ band strength ratio, where the numerator traces aliphatic and the denominator aromatic carbonaceous material (PAHs). Apart from some `hot' pixels, there is a clear demarcation between the dense and diffuse region, with the ratio systematically higher in the former.

Plotting I\textquotesingle$_{\rm 3.4}$ versus I\textquotesingle$_{\rm 3.3}$ in the right panel of Figure~\ref{fig:aliphatics} reveals a bifurcation that matches with what is observed in the left panel. This is best seen when color coding each point according their zone assignment from Figure~\ref{fig:hst}, which is based on hierarchical clustering of the Spitzer spectra \citep[][]{2018ApJ...858...67B}. Points associated with the diffuse (yellow) and dense medium (green) make up two distinct trends. Similar behavior has been observed for the 11.2 versus 8.6~\textmu m PAH band strength \citep[][]{2015ApJ...806..121B, 2016ApJ...832...51B, 2018ApJ...858...67B} for NGC7023, which was related to an evolution in PAH charge. \citet{2025arXiv251006167K} observed branching in some of their correlation plots involving the 8.6~\textmu m PAH band in JWST observations of the Orion Bar. These authors conclude that only a subset of the interstellar PAH family contribute to the 8.6~\textmu m PAH band and, subsequently, attribute them to changes in PAH size.

As far as I\textquotesingle$_{\rm 3.4}$/I\textquotesingle$_{\rm 3.3}$ is concerned, its evolution across the northwest PDR of NGC7023 is consistent with an interpretation that aliphatic hydrogen side groups have a better chance of surviving in more shielded, denser regions but, once exposed, are rapidly photo-dissociated from the parent PAH. This includes both superhydrogenated and  methylated PAHs, where the former have been shown to produce a better match with observations \citep[][]{1996ApJ...472L.127B, 2013ApJS..205....8S, 2020ApJS..247....1Y}. Experiments suggest that PAHs with different side groups are released from ices at the PDR-front, where their production and residence time in the ice has promoted superhydrogenation and the addition of  side groups \citep[][]{1995A&A...295..479M, 1999Sci...283.1135B, 2002ApJ...576.1115B}

Following \citet{2023ApJS..268...50Y}, the ratio of the number of aliphatic ($N_{\rm H,ali}$) relative to aromatic ($N_{\rm H,aro}$) hydrogens in the dense and diffuse medium is found to be $[N_{\rm H,ali}/N_{\rm H,aro}]_{\rm dense}$ = 0.42\textpm0.01 and $[N_{\rm H,ali}/N_{\rm H,aro}]_{\rm diffuse}$ = 0.10\textpm0.01, respectively. The latter value is somewhat consistent with that found by \citeauthor{2025A&A...700A.158M} (\citeyear{2025A&A...700A.158M}; $0.062\lesssim N_{\rm H,ali}/N_{\rm H,aro}\lesssim0.23$) from JWST observations centered on the peak of the PAH emission at the PDR-front. Though, the previous value for the dense medium is higher by a factor of two. This difference is attributed, in part, to the FOV considered here being much larger (50\textquotedbl.4$\times$50\textquotedbl.4 versus $\sim$3\textquotedbl.3$\times$21\textquotedbl) and extending further into the more shielded, molecular regions. Moreover, the entire JWST FOV is covered by a mere $\sim$0.5$\times$3 SPHEREx+Spitzer pixels. In addition, because of the two orders of magnitude difference in spectral resolution, \citet{2025A&A...700A.158M} are able to isolate the 3.3 and 3.4~\textmu m features from each other and any other components far more carefully, which impacts the determined band strengths.

\subsection{The 11.2/3.3 \texorpdfstring{\textmu}{u}m Band Strength Ratio}
\label{subsec:112_33}

\begin{figure*}
    \includegraphics[width=\textwidth]{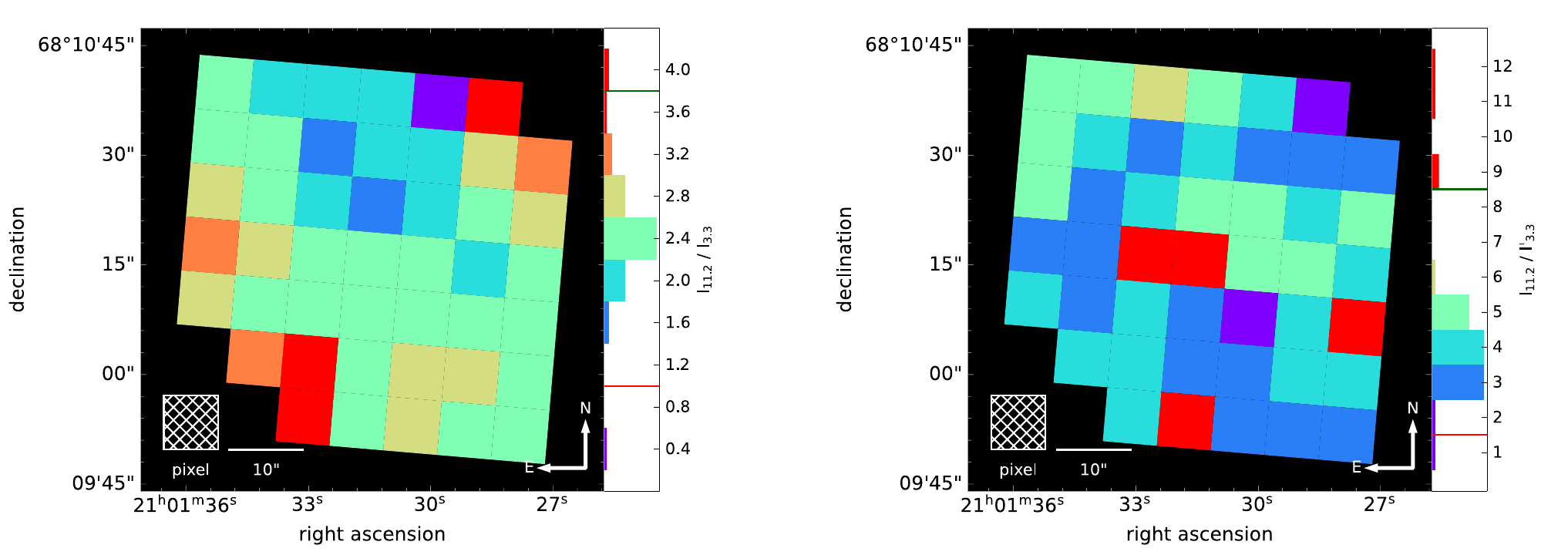}
    \caption{Maps of the northwest PDR in NGC7023 in the 11.2/3.3~\textmu m PAH band strength ratio. The 3.3~\textmu m band strength determined through direct integration (I$_{\rm 3.3}$) and spectro-photometric modeling (I\textquotesingle$_{\rm 3.3}$) are presented in the left and right panel, respectively. The color scales use a 4\% cutoff on the high and a 1\% cutoff on the low end, indicated by the green and red lines in the color bar, respectively. This suppresses outliers. See Section~\ref{subsec:112_33} for details}
    \label{fig:112_33}
\end{figure*}

Figure~\ref{fig:112_33} presents maps of the 11.2/3.3~\textmu m PAH band strength ratio. The figure's left panel uses the 3.3~\textmu m band strength determined through direct integration (I$_{\rm 3.3}$; Section~\ref{subsec:direct}) while the right panel that derived from the spectro-photometric modeling (I\textquotesingle$_{\rm 3.3}$; Section~\ref{subsec:model}).

The 11.2/3.3~\textmu m PAH band strength ratio has been shown to be an important tracer for PAH size \citep[e.g.,][]{1989ApJS...71..733A, 2020MNRAS.494..642M, 2021ApJ...917....3D, 2023MNRAS.524.3429M}. At face value, the survivability of smaller PAHs is expected to decrease when moving from the shielded dense medium into the exposed diffuse medium, mirroring the behavior of I\textquotesingle$_{3.4}$/I\textquotesingle$_{3.3}$ in Figure~\ref{fig:aliphatics}. Indeed, the left panel of Figure~\ref{fig:112_33} shows a clear demarcation with smaller PAHs, i.e., smaller 11.2/3.3 ratios, more prominent across the dense medium. This is consistent with results from \citet{2016A&A...590A..26C} that are based on a combination of Stratospheric Observatory For Infrared Astronomy \citep[SOFIA;][]{2012ApJ...749L..17Y} FLITECAM and FORCAST observations.

However, the right panel of Figure~\ref{fig:112_33}, which replaces I$_{\rm 3.3}$ with I\textquotesingle$_{\rm 3.3}$, paints a different, far more chaotic picture. Gone are the clear demarcation and the smooth transition between the dense and diffuse medium when crossing the PDR-front. This difference could be due to a reduction in signal quality, but the dynamic ranges, 6.3 versus 4, remain comparable. Arguably, limitations of the spectro-photometric model used to isolate the 3.3~\textmu m feature from the 3.4~\textmu m sub-feature could be the cause. For example, the model described in Appendix~\ref{app:model} combines the 3.4~\textmu m feature and underlying plateau that is shared with the 3.3~\textmu m feature into a single Gaussian. Nevertheless, the 3.4/3.3~\textmu m results do demonstrate merit for this approach. 

Recent work \citep[][]{2024MNRAS.535.3239E} has shown that the distinction between aliphatic and aromatic C-H stretch features is not as clear cut as previously thought. This, in combination with the above, could also be a contributing factor for the observed differences when using I\textquotesingle$_{\rm 3.3}$ in the ratio.

\subsection{The PAH Size Evolution across NGC7023}
\label{subsec:pahdb}

\begin{figure*}
   \includegraphics[width=\textwidth]{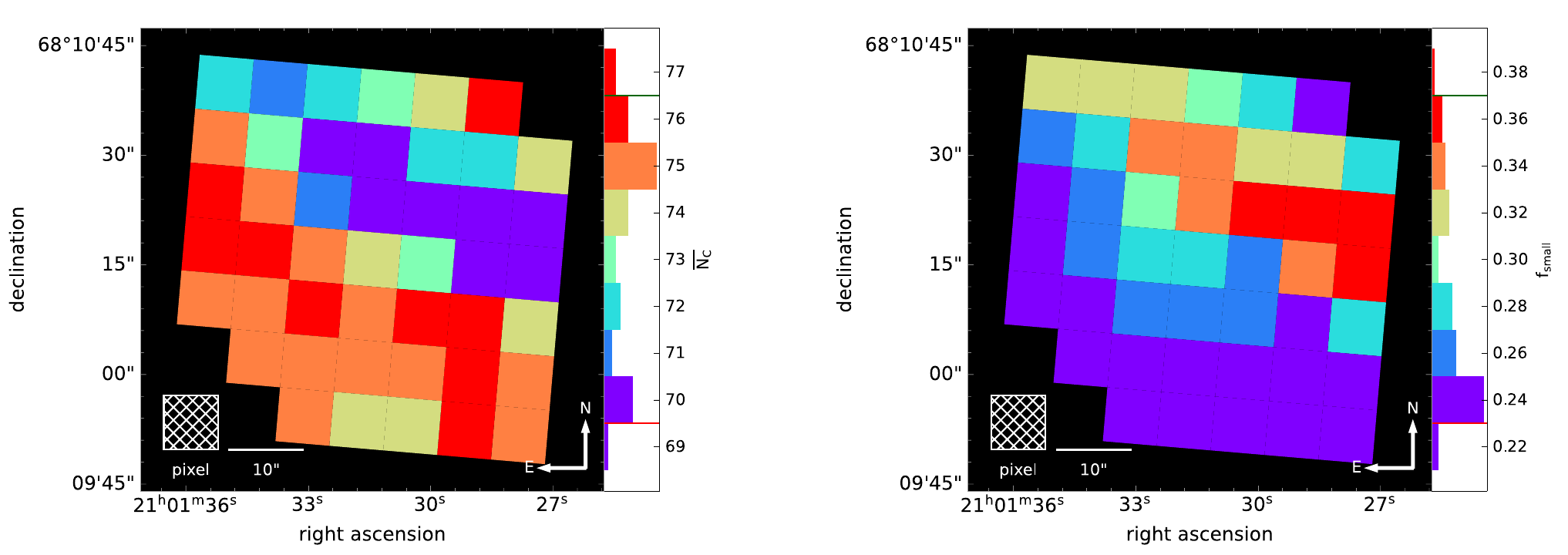}
   \caption{Maps of the northwest PDR in NGC7023 showing the average number of carbon atoms ($\overline{\rm N_{C}}$; left) and small PAH fraction (f$_{\rm small}$; N$_{\rm C}\le50$; right) determined from a Monte Carlo PAHdb-fit to each pixel's corresponding Spitzer segment isolated PAH spectrum. The color scales use a 4\% cutoff on the high and a 1\% cutoff on the low end, indicated by the green and red lines in the color bar, respectively. This suppresses outliers. See Section~\ref{subsec:pahdb} for details.}
   \label{fig:mcfit}
\end{figure*}

To obtain a quantitative measure for the evolution of PAH size across NGC7023, the  NASA Ames PAH IR Spectroscopic Database \citep[PAHdb hereafter;][]{2010ApJS..189..341B, 2014ApJSS..211....8B, 2018ApJS..234...32B, 2020ApJS..251...22M} is employed to fit the 5.2--15~\textmu m isolated, extinction corrected Spitzer PAH segment of the spectra. This approach is largely analogous to that used by \citet{2013ApJ...769..117B, 2015ApJ...806..121B, 2016ApJ...832...51B, 2018ApJ...858...67B}, the main differences being: \textit{1.} the use of a much more recent version \citep[4.00-$\alpha$;][]{2025ApJ...979...90M} of PAHdb's library of computed PAH spectra, \textit{2.} exclusion of nitrogen containing PAHs \citep[PANHs; per][]{2024ApJ...968..128R}, \textit{3.} \emph{no} application of an emission redshift \citep[per][]{2018JChPh.149m4302M}, \textit{4.} incorporating an excitation energy of 7~eV, and \textit{5.} having Gaussian line profiles with a fixed full-width at half maximum (FWHM) of 15~cm$^{\rm -1}$. This matches the optimal PAHdb modeling configuration from \citet{2025ApJ...979...90M}. A Monte Carlo approach is taken where the spectra are varied within their uncertainties 1024 times and fitted to obtain a statistical measure, i.e., mean and standard deviation, for derived parameters\footnote[2]{These do not capture uncertainties associated with the PAHdb spectra or the emission modeling, neither statistical nor systematic.}. The interaction with the library, emission modeling, and fitting are done with PAHdb's suite of Python tools\footnote[3]{\href{http://github.com/PAHdb/AmesPAHdbPythonSuite}{github.com/PAHdb/AmesPAHdbPythonSuite}} \citep[][]{zenodo.17428588}.

Figure~\ref{fig:mcfit} presents maps of the determined average number of carbon atoms $\overline{\rm N_{C}}$ and small PAH fraction f$_{\rm small}$ (N$_{\rm C}\le50$). Despite their limited span, especially for $\overline{\rm N_{C}}$ ($\sim$70-76), both maps pick up the demarcation between the dense and diffuse medium toward the north and south of the FOV, respectively. As expected, $\overline{\rm N_{C}}$ increases and f$_{\rm small}$ drops across the PDR and into the diffuse medium. Compared to \citet{2016A&A...590A..26C}, who find on average N$_{\rm C}\simeq70$ in the diffuse medium and $\sim$50 at the PDR surface, the upper boundary in the present data is slightly higher (76) and $\overline{\rm N_{C}}$ does not drop below 70.

\begin{figure*}
   \includegraphics[width=\textwidth]{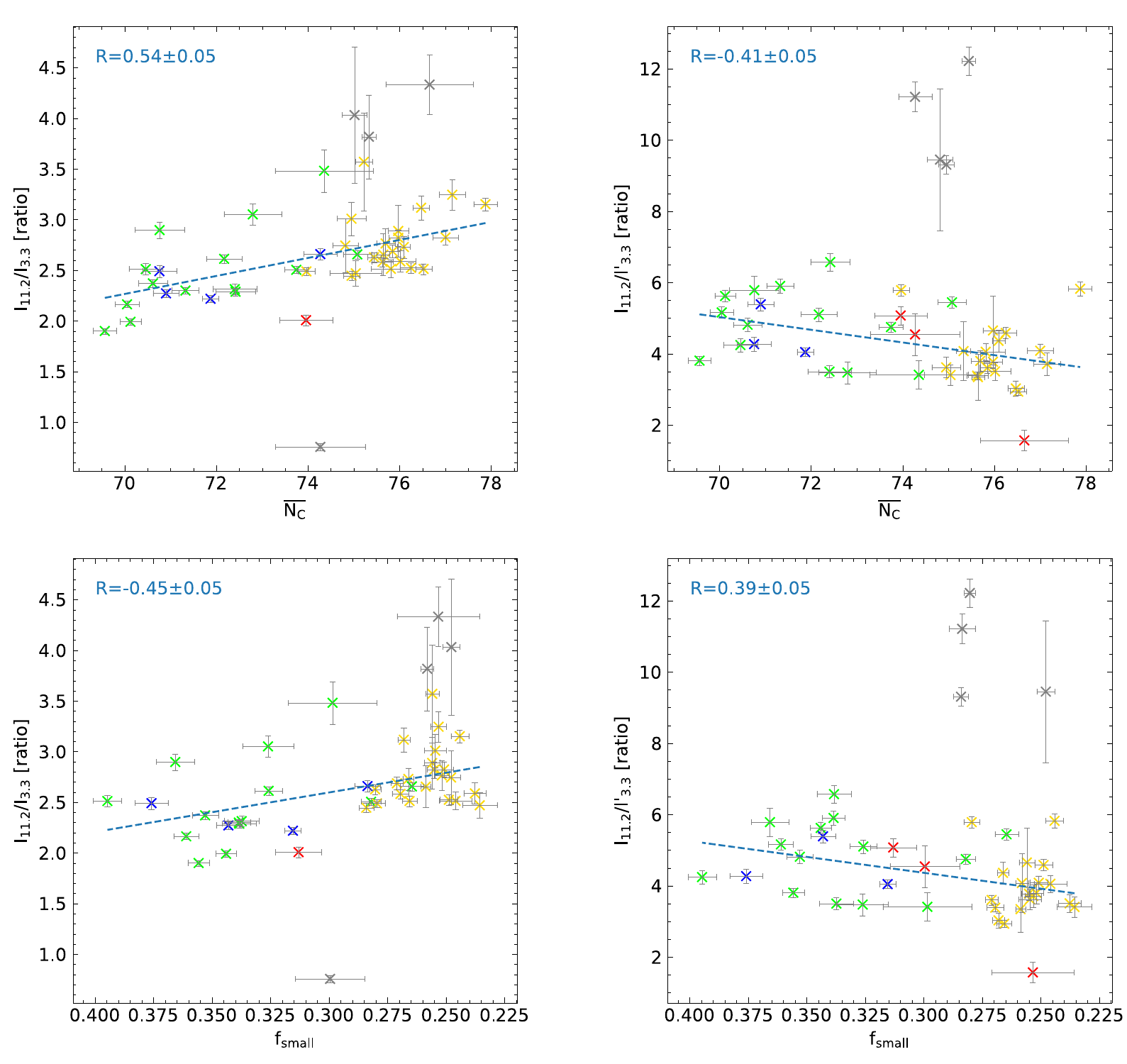}
   \caption{PAH band strength ratio and PAHdb-derived size correlations. The top two panels present the 11.2/3.3~\textmu m PAH band strength ratio against the PAHdb-determined average number of carbon atoms ($\overline{\rm N_{C}}$) whilst the bottom two panels that versus the small PAH fraction f$_{\rm small}$ (N$_{\rm C}\le50$). Panels to the left and right use the 3.3~\textmu m PAH band strength determined through direct integration (I$_{\rm 3.3}$) and spectro-photometric modeling (I\textquotesingle$_{\rm 3.3}$), respectively. Points are color coded according their zone as assigned in Figure~\ref{fig:hst}. Clipped points are shown in gray (3$\sigma$). Uncertainties have been propagated, indicated as gray error bars, and data points with a SNR less than three have been discarded. Given is the weighted Pearson correlation coefficient R (blue) for a straight line (dashed). See Section~\ref{subsec:pahdb} for details.}  
   \label{fig:size}
\end{figure*}

With the 11.2/3.3~\textmu m PAH band strength ratio and the quantitative measures for PAH size from PAHdb, correlating one with the other has the potential to provide a direct calibrator for PAH size. This is analogous to what has been done for PAH charge, where the 6.2/11.2~\textmu m PAH band strength ratio is correlated with the PAHdb-derived PAH cation/neutral ratio, which, in turn, is connected to the PAH ionization parameter \citep[$\gamma$; e.g.,][]{2008ApJ...679..310G, 2018ApJ...858...67B, 2022ApJ...931...38M}.

The top two panels of Figure~\ref{fig:size} present 11.2/3.3 versus $\overline{\rm N_{C}}$. In the left and right panel the 3.3~\textmu m PAH band strength determined through direct integration (I$_{\rm 3.3}$) and spectro-photometric modeling (I\textquotesingle$_{\rm 3.3}$) are shown, respectively. With a sigma-clip of 3$\sigma$ and disregarding points with a signal-to-noise ratio (SNR) of less than 3, a tentative, slightly positive trend appears (R=0.54\textpm0.05) for the former. However, use of I\textquotesingle$_{3.3}$ reveals a less convincing negative correlation (R=-0.41\textpm0.05). This behavior is driven by I\textquotesingle$_{3.3}$ and reflects that seen in the right panel of Figure~\ref{fig:112_33}. Moreover, the range in $\overline{\rm N_{C}}$ is very limited, spanning only $\sim$6 carbon atoms, despite I$_{\rm 11.2}$/I\textquotesingle$_{\rm 3.3}$ covering much of the observed astronomical variation \citep[$\sim$4-7; e.g.,][]{2020MNRAS.494..642M}. In addition, careful analysis of JWST observations of the Orion Bar with PAHdb are able to recover the expected trend, albeit after averaging a large number of points \citep[][]{Maragkoudakis2025b}. Section~\ref{subsec:112_33} provides possible explanations for the discrepancy seen here.

In the bottom two panels of Figure~\ref{fig:size} 11.2/3.3 versus f$_{\rm small}$ is considered. As with $\overline{\rm N_{C}}$, f$_{\rm small}$ shows a tentative trend when using I$_{3.3}$ (R=-0.45\textpm0.05) and a less convincing one with I\textquotesingle$_{3.3}$ (R=0.39\textpm0.05). The range in f$_{\rm small}$ is relatively small, only spanning $\sim$0.14. In contrast, the range in the PAHdb-derived PAH cation fraction is almost three times larger ($\sim$0.41) and remains above $\sim$0.60 throughout the FOV.

On top of the possible reasons given in Section~\ref{subsec:112_33}, it could be that the intrinsic intensity of the C-H stretches and bends is a contributing factor for not observing stronger correlations. That is, C-H stretches and bends are significantly stronger in neutral PAHs than their ionized counterparts. With larger PAHs more readily being ionized, and since the environment is highly ionized, the 11.2/3.3~\textmu m PAH band strength ratio might not fully capture the subpopulation of large ionized PAHs.

Another possibility is that the ongoing photo-chemistry does not allow for a straightforward interpretation of PAH size, certainly not as easily as has been the case for PAH charge. Perhaps this is related to how the assorted population of freshly minted PAHs sublimated off icy grains entering the diffuse medium at the PDR-front is steadily evolving toward one of mainly fullerenes when approaching HD200775 \citep[][]{2012PNAS..109..401B, 2013A&A...550L...4B}. A larger, more varied sample of sizable extended objects of combined SPHEREx and Spitzer spectral mapping observations will be needed to provide stronger evidence for the reported trends. 

On the database fitting side, directly including the 3.3~\textmu m PAH feature could provide an additional constraint that might break, if present, any size degeneracy and alter the trends favorably. However, this is not straightforward due to the spectral complexity of the 3-5~\textmu m region \citep[see e.g.,][]{2023ApJ...959...74B, 2025A&A...701A.111V}. Moreover, the picture painted thus far already aligns with that of rapid photo-alteration of the population as soon as the PAHs leave the dense medium and larger PAHs becoming more prominent members when moving through the diffuse medium closer toward HD200775.

\section{Conclusions}
\label{sec:conclusions}

The following conclusions are drawn:\\

\begin{enumerate}[nosep, leftmargin=*]

\item Observations by the SpectroPhotometer for the History of the Universe, Epoch of Reionization, and Ices Explorer (SPHEREx) of the Galactic interstellar medium enable the study of the aromatic and aliphatic evolution across large extended objects by providing extraordinary access to the 3.3 polycyclic aromatic hydrocarbon (PAH) and 3.4~\textmu m aliphatic features, respectively.

\item The 3.3 and 3.4~\textmu m emission in the northwest PDR of NGC7023 peaks at the PDR-front, both dropping off when moving into the diffuse medium in the direction of the irradiating star, HD200775, located toward the south-east, and when moving into the direction of the embedded young stellar object in the northwestern corner of the field-of-view (FOV).

\item The 3.4/3.3~\textmu m band strength ratio in NGC7023 shows a demarcation that matches the large-scale morphology of the region, which, in turn, is consistent with an interpretation that PAH superhydrogenation and aliphatic side groups are able to survive in more shielded regions but get quickly removed when entering the unshielded, diffuse medium. This demarcation shows up as two distinct trends when plotting the 3.4 versus the 3.3~\textmu m band strength, one associated with the dense medium and the other with the diffuse medium; $[N_{\rm H,ali}/N_{\rm H,aro}]_{\rm dense}$ = 0.42\textpm0.01 and $[N_{\rm H,ali}/N_{\rm H,aro}]_{\rm diffuse}$ = 0.01\textpm0.01.

\item In conjunction with matching Spitzer spectral map observations, SPHEREx enables the study of the PAH size evolution across large extended objects by providing access to the 3.3~\textmu m PAH band that, when ratioed against the 11.2~\textmu m PAH band, is considered the primary observational tracer for PAH size.

\item The 11.2/3.3~\textmu m PAH band strength ratio in NGC7023 also follows the large-scale morphology of the region, with the same demarcation. Here, this is consistent with an interpretation that the smaller PAHs in the population are quickly whittled away when entering the more exposed, diffuse medium. However, this is only the case when considering the 3.3--3.4~\textmu m complex and not so when it is decomposed into a separate 3.3 and 3.4~\textmu m sub-feature.

\item NASA Ames PAH IR Spectroscopic Database Monte Carlo fits to the isolated $\sim$5--15~\textmu m Spitzer PAH spectra to retrieve the average number of carbon atoms ($\overline{\rm N_{C}}$) and small PAH fraction ($f_{\rm small}$; N$_{\rm C}\le50$) are able to pick up the demarcation between the dense and diffuse medium, with $\overline{\rm N_{C}}$ increasing and $f_{\rm small}$ decreasing toward HD200775, respectively. This is despite little variation across the FOV; $70\lesssim\overline{\rm N_{C}}\lesssim76$ and $0.24\lesssim\text{f}_{\rm small}\lesssim0.36$.

\item The 11.2/3.3 PAH band strength ratio versus $\overline{\rm N_{C}}$ and $f_{\rm small}$, show tentative correlations (R=0.54\textpm0.05 and -0.45\textpm0.05, respectively). Once again, this is only the case when considering the 3.3--3.4~\textmu m complex. In the case of the 3.3~\textmu m PAH band strength from the decomposed complex, the correlations are flipped and less tight with R=-0.41\textpm0.05 and 0.39\textpm0.05 for $\overline{\rm N_{C}}$ and $f_{\rm small}$, respectively.

\item Studying a sample of various large(r) extended objects with varying astrophysical and astrochemical conditions is needed to put the reported trends on a stronger footing, tighten the correlations, and turn the 11.2/3.3~\textmu m into a quantitative calibrator for PAH size. Moreover, it will expose the reason for the discrepancy between the results when considering the 3.3--3.4~\textmu m complex as a whole and the decomposed 3.3~\textmu m sub-feature on its own.

\end{enumerate}

\begin{acknowledgments}

This publication makes use of data products from the Spectro-Photometer for the History of the Universe, Epoch of Reionization and Ices Explorer (SPHEREx), which is a joint project of the Jet Propulsion Laboratory and the California Institute of Technology, and is funded by the National Aeronautics and Space Administration (NASA). In addition, this work is based in part on observations made with the Spitzer Space Telescope, which was operated by the Jet Propulsion Laboratory, California Institute of Technology under a contract with NASA. C.B. is grateful for an appointment at NASA Ames Research Center through the San Jos\'e State University Research Foundation (80NSSC22M0107). L.J.A., A.M., and J.D.B. are thankful for an appointment at NASA Ames Research Center through the Bay Area Environmental Research Institute (80NSSC23M0028). R.C.F. acknowledges funding support from NASA Grant 80NSSC24M0132. C.B., A.M., L.J.A., J.D.B., P.T., V.J.E., and R.C.F. acknowledge support from the Internal Scientist Funding Model (ISFM) Laboratory Astrophysics Directed Work Package at NASA Ames.

\end{acknowledgments}

\begin{contribution}

C.B. was responsible for conceptualization, data curation, formal analysis, investigation, methodology, project administration, validation, visualization, and writing; both the original draft and during review \& editing. A.M. aided with the methodology. A.M., L.J.A., J.D.B., P.T., V.J.E., and R.C.F. were responsible for validation and writing; both the original draft as well as during review \& editing.

\end{contribution}

\facilities{SPHEREx, Spitzer(IRS)}

\software{AmesPAHdbPythonSuite \citep[][]{zenodo.17428588}, astropy \citep[][]{2013A&A...558A..33A, 2018AJ....156..123A, 2022ApJ...935..167A}, photutils \citep[][]{larry_bradley_2025_14889440}, pyphot \citep[][]{zenodo14712174}}

\bibliography{aamnem99,bibliography}{}
\bibliographystyle{aasjournalv7}

\appendix
\restartappendixnumbering

\section{Extinction Correction}
\label{app:extinction}

\begin{figure}
    \centering
    \includegraphics[width=\columnwidth]{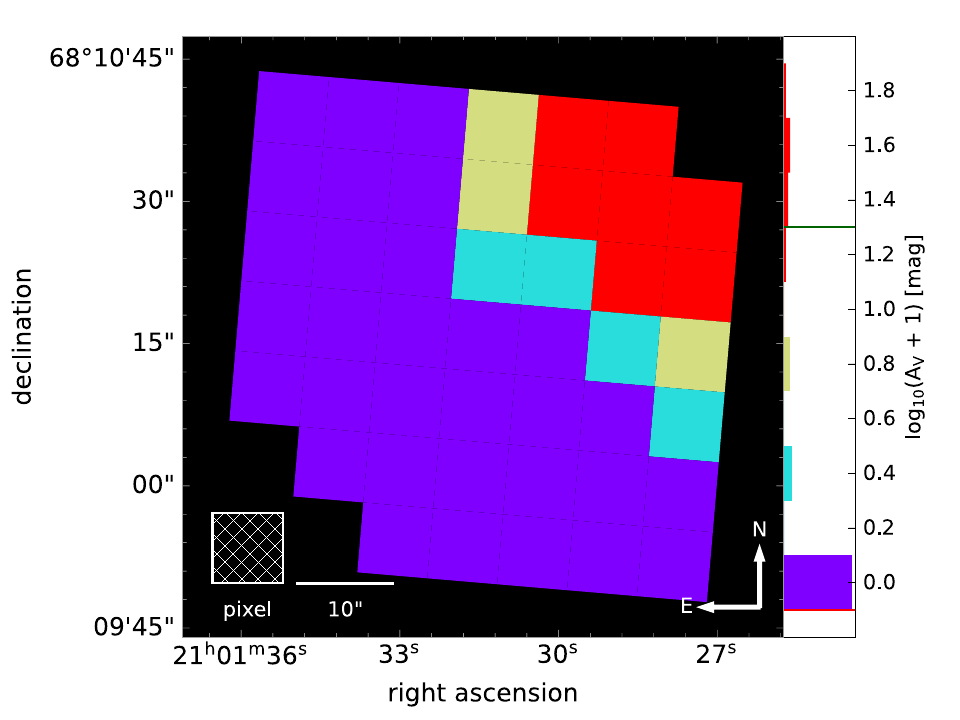}
    \caption{Visual extinction (A$_{\rm V}$) across the northwest PDR in NGC7023. The color scale uses a 4\% cutoff on the high and a 1\% cutoff on the low end, indicated by the green and red lines in the color bar, respectively. This suppresses outliers. See Appendix~\ref{app:extinction} for details.}
    \label{fig:extinction}
\end{figure}

Spectra have been corrected for extinction. The visual extinction (A$_{\rm V}$) determined by \citet{2018ApJ...858...67B} is taken, which was derived using the Galactic center extinction curve from \citet{2006ApJ...637..774C}. See \citet{2018ApJ...858...67B} for additional detail. Figure~\ref{fig:extinction} presents a map of the visual extinction, resampled onto the SPHEREx FOV.

\section{Splicing}
\label{app:splicing}

\begin{figure}
    \centering
    \includegraphics[width=\columnwidth]{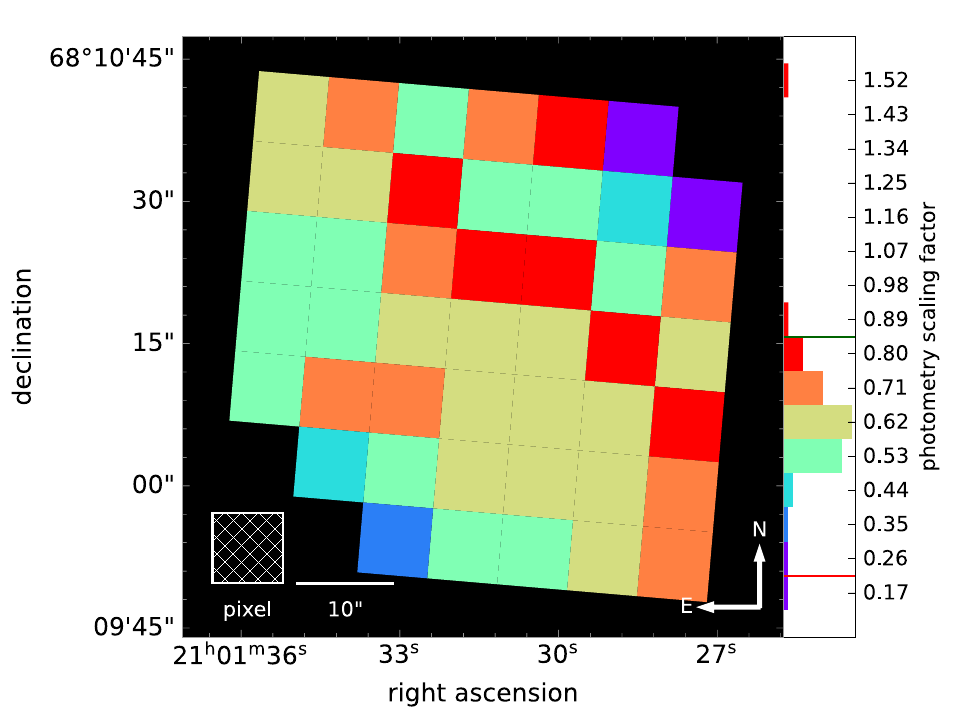}
    \caption{SPHEREx segment scaling factors across the northwest PDR in NGC7023 determined from predicted and actual IRAC channel 1 and 4 photometry. The color scale uses a 4\% cutoff on the high and a 1\% cutoff on the low end, indicated by the green and red lines in the color bar, respectively. This suppresses outliers. See Appendix~\ref{app:splicing} for details.}
    \label{fig:factors}
\end{figure}

As there is no overlap between the SPHEREx and Spitzer segments, they cannot be directly spliced together. Therefore, the predicted and observed IRAC photometry are used to scale the SPHEREx segment per:
\begin{equation}
    \label{eq:irac}
    \frac{SF\cdot\text{IRAC1}_{\rm predicted}}{\text{IRAC4}_{\rm predicted}} = \frac{\text{IRAC1}_{\rm observed}}{\text{IRAC4}_{\rm observed}}\ .
\end{equation}

Figure~\ref{fig:factors} presents a map of the derived scaling factors $SF$. Since $SF$ scales the relative photometric bands, no assessment is made on the absolute flux calibration of either the SPHEREx or Spitzer segments. Given that this is a single multiplicative factor, scaling factors are applied to the SPHEREx band strength measurements and not to the segment as a whole before the analysis.

\section{Spectro-photometric Model}
\label{app:model}

The model used in Section~\ref{subsec:model} to reproduce the SPHEREx bandpass-integrated spectro-photometry solves for:
\begin{equation}
     \label{eq:match}
     F_{\rm\lambda,SPHEREx} \Delta\lambda_{\rm SPHEREx} -  \int\limits_{\Delta\lambda_{\rm SPHEREx}}F_{\lambda,\rm model}(\lambda)\mathrm{d
    }\lambda\ = 0,
\end{equation}
where $F_{\rm\lambda,SPHEREx}$ is the photometric data point in W~cm$^{\rm -2}$~\textmu m$^{\rm -1}$, $\Delta\lambda_{\rm SPHEREx}$ the bandpass in \textmu m, and $F_{\lambda,\rm model}(\lambda)$ the model spectrum with units of W~cm$^{\rm -2}$~\textmu m$^{\rm -1}$. The integration of $F_{\lambda,\rm model}(\lambda)$ is done over the bandpass $\Delta\lambda_{\rm SPHEREx}$.

$F_{\lambda,\rm model}(\lambda)$ consists of two blended Gaussian profiles that describe the 3.29 and 3.4~\textmu m features, respectively, and a straight baseline:

\begin{equation}
    \label{eq:model}
    F_{\lambda,\rm model}(\lambda) = m\lambda + b +\sum\limits_{i=1}^{2} G_{\lambda,i}(\lambda)\ ,
\end{equation}
where $m$ and $b$ are the slope and offset of the straight baseline, respectively. $G_{\lambda,i}(\lambda)$ represent the Gaussian profile:

\begin{equation}
   \label{eq:gaussian}
   G_{\lambda,i}(\lambda) = A_{i}e^{-\frac{(\lambda - \lambda_{0,i})^{2}}{2\sigma_{i}^{2}}},
\end{equation}
where $A_{i}$, $\lambda_{0,i}$, and $\sigma_{i}$ stand for the center flux in W~cm$^{\rm -2}$~\textmu m$^{\rm -1}$, center wavelength in \textmu m, and width in \textmu m, respectively. The band strength $G_{\lambda,i}$ is then determined as
\begin{equation}
   \label{eq:power}
   G_{\lambda,i} = A_{i}\sigma_{i}\cdot\sqrt{2\pi}\ .
\end{equation}

$A_{i}$ is forced strictly positive, line centers $\lambda_{0,i}$ are set to 3.279 and 3.40~\textmu m, and widths to 0.0247 and 0.0807~\textmu m for the 3.3 and 3.4~\textmu m components, respectively. Line centers are allowed to vary within 0.04~\textmu m and the widths $\sigma_{i}$ between 0.02 and 0.3 and 0.07 and 0.09~\textmu m for the 3.3 and 3.4~\textmu m components, respectively. Uncertainties are propagated. 

\end{document}